\documentclass[final,5p,times,twocolumn]{elsarticle}

\usepackage{amssymb}
\usepackage{amsmath}
\usepackage{amsfonts}
\journal{Physica B}

\begin{document}

\begin{frontmatter}



\title{First principles determination of the model parameters in $\kappa$-(ET)$_2$Cu$_2$(CN)$_3$}


\author[label1]{Harald O. Jeschke}
\author[label2]{Hem C. Kandpal}
\author[label1]{Ingo Opahle}
\author[label1]{Yu-Zhong Zhang}
\author[label1]{Roser Valent{\'\i}}

\address[label1]{Institut f\"ur Theoretische Physik, Goethe-Universit\"at
Frankfurt, Max-von-Laue-Stra{\ss}e 1, 60438 Frankfurt am Main, Germany}
\address[label2]{IFW Dresden, P.O. Box 270016, D-01171 Dresden, Germany}

\begin{abstract}
We present a detailed study of the derivation of the Hubbard model parameters
for $\kappa$-(ET)$_2$Cu$_2$(CN)$_3$ in the framework of
 {\it ab initio} Density Functional Theory. We show that calculations
with  different (i) wavefunction basis, (ii) exchange correlation functionals
and (iii) tight-binding models provide a reliable benchmark for the parameter values.
We compare our results with available extended H\"uckel molecular orbital calculations
and discuss its implications for the description of the properties of  $\kappa$-(ET)$_2$Cu$_2$(CN)$_3$.
The electronic properties of $\kappa$-(ET)$_2$Cu(SCN)$_2$ are also briefly discussed.
\end{abstract}


\end{frontmatter}


\section{Introduction}
\label{}

The search for possible candidates of a quantum spin liquid state has
been one of the central issues in condensed matter physics since it
was proposed several decades ago~\cite{Anderson73}. In the last years
$\kappa$-(ET)$_2$Cu$_2$(CN)$_3$ has attracted much attention due to its
unique realization of a pressure-induced Mott transition from a spin
liquid state~\cite{Shimizu03,YamashitaS08,YamashitaM08}
 to a metallic or even superconducting state at low
temperatures.

A large amount of theoretical investigations on the
 one-band half-filled Hubbard model
on the anisotropic triangular lattice have been carried out in order to
understand the mechanism of the phase transition. Irrespective of
the current debate on the phase diagram, such as presence or absence of
spin-liquid or superconducting state, derived from this simplified
model by various methods, like, path integral renormalization group
approach~\cite{Mizusaki06},  variational  Monte Carlo
method~\cite{Watanabe08,Tocchio09}, 
cluster extension of dynamical mean-field theory~\cite{Kyung06}, 
or exact diagonalization~\cite{Clay08} a question
of fundamental importance is whether the existing model parameters
obtained from extended H\"uckel molecular orbital calculations~\cite{Komatsu96} are
reliable or not, as they are the basis of all the mentioned theoretical
work.

Recently we showed by full potential ab initio Density Functional Theory (DFT)
calculations that
the model parameters for $\kappa$-(ET)$_2$Cu$_2$(CN)$_3$ have to be
qualitatively revised from $t'/t$=1.06~\cite{Komatsu96} to $t'/t$=0.83$\pm$0.08~\cite{Kandpal09},
which casts doubts on the validity of the description of the system 
 as coupled quasi-one-dimensional chains. Similar results were obtained
by the authors of Ref.~\cite{Nakamura09}. 
In the present paper, we  discuss  the
importance of including in the results for $t'/t$ 
 the effects of considering different wavefunction bases as well as
 exchange correlation functionals within DFT. Furthermore, 
we show that alternative fitting procedures provide slightly different
$t'/t$ ratios.  All these considerations lead to a statement of the $t'/t$ ratios
within a benchmark. Finally we also discuss the effects of longer ranged hoppings.
These considerations are important for all  $\kappa$-(ET)$_2$X charge transfer salts and
we present here also details of the bandstructure calculations and parameter determination
for X=Cu(SCN)$_2$.

\section{Crystal structure and bandstructure calculations}

\begin{figure}
\includegraphics[width=0.3\textwidth]{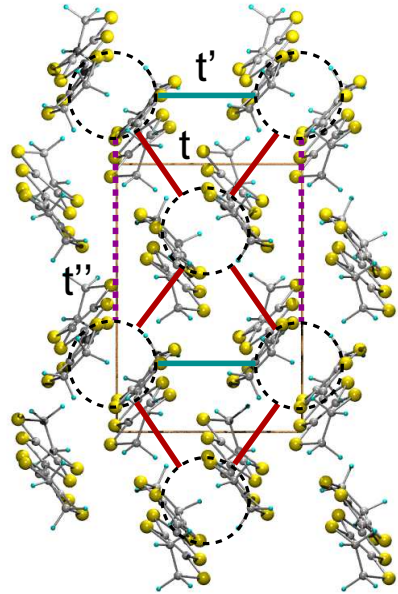}
\caption{Structure of the $\kappa$ BEDT-TTF arrangement for the example of $\kappa$-(BEDT-TTF)$_2$X with   X=Cu$_2$(CN)$_3$. $t$ and $t'$ form a frustrated triangular lattice.}
\label{cn_structure}
\end{figure}

The crystal structure of $\kappa$-(ET)$_2$Cu$_2$(CN)$_3$ 
consists of  alternate stackings  of pairs of ET = BEDT-TTF  molecule 
donors (see Fig.~\ref{cn_structure})  separated by Cu$_2$(CN)$_3$ acceptor layers. 
Each BEDT-TTF molecule consists of 10 carbon, 8 sulfur 
and 8 hydrogen atoms linked by covalent bonds. 
A common feature of $\kappa$-(ET)$_2$Cu$_2$(CN)$_3$ and other related 
BEDT-TTF based salts is the presence of conformational disorder. 
The two terminal ethylene groups in the 
BEDT-TTF molecule are twisted out of the molecular plane, and therefore this 
molecule has either an eclipsed or staggered conformation. The salt 
exhibits disorder at room temperature due to the random occurence of
both conformations. The $\kappa$-(ET)$_2$Cu$_2$(CN)$_3$ salt crystallizes in the 
monoclinic space group $P2_1/c$. The inversion centre symmetry in $P2_1/c$ 
is guaranteed by the equal occupation of the two conformations.
At low temperatures the staggered conformation dominates over the
eclipsed conformation, therefore the staggered conformation was used in the
calculations. The symmetry is then reduced to $Pc$ with 120 atoms in the 
unit cell.

Moreover, since the hydrogen positions were not available from 
experiment~\cite{Geiser91} we introduced them in the structure
and relaxed their positions by performing Car-Parrinello projector-augmented wave
molecular dynamics (CPMD) calculations~\cite{Car85,Parrinello80,Bloechl94}. For these calculations we
considered a (4 $\times$ 4 $\times$ 4) {\bf k}-mesh in the Brillouin zone
 and plane wave cutoffs of 60 and 240 Ry for the wave function and the charge density, respectively.

The electronic structure was analyzed by considering two different all-electron 
codes, namely the the Full Potential Local Orbital method 
(FPLO version 8.50-32)~\cite{FPLO,Opahle09} 
and the Full Potential Linearized Augmented Plane Wave (FPLAPW) method 
as implemented in  Wien2k~\cite{WIEN}. We also performed comparative 
calculations with different exchange correlation functionals, the 
Local Density Approximation (LDA) in the parameterization of Perdew and 
Wang~\cite{Per92} and the Generalized Gradient Approximation (GGA) in 
the Perdew-Burke-Ernzerhof parameterization~\cite{PBE96}.
The convergency of the band structure with respect to the Brillouin zone
integrations was checked with a series of calculations with up to 350
{\bf k}-points in the full Brillouin zone.
In the FPLAPW calculations a plane wave cut-off 
$R_{MT} \times k_{\rm max} = 3.37$ was used and the  muffin-tin-radii 
for Cu, N, C, H, and S were chosen as 2.37, 1.06, 1.16/1.06, 0.92, and 
2.00 a.u., respectively.

\begin{figure}
\includegraphics[width=0.47\textwidth]{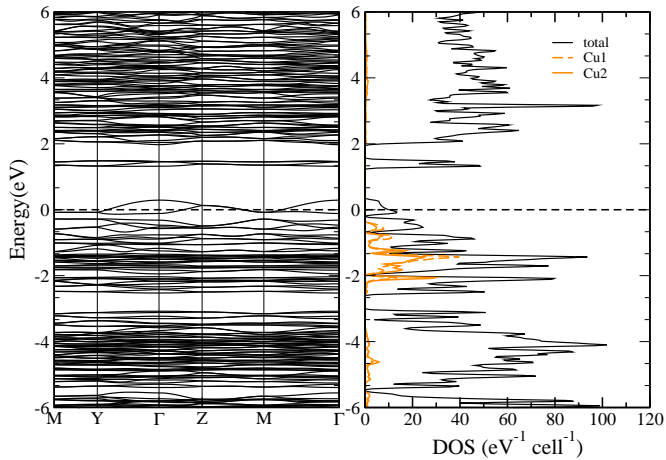}
\caption{Band structure and density of states of $\kappa$-(BEDT-TTF)$_2$X with   X=Cu$_2$(CN)$_3$, calculated with FPLAPW. The contribution of Cu to the DOS is highlighted. The Cu $3d$ states are fully occupied, indicating that Cu is in a nonmagnetic Cu$^{+1}$ oxidation state.
}
\label{FIG:bsdos}
\end{figure}

Despite the complexity of the crystal structure, the electronic structure of
$\kappa$-(BEDT-TTF)$_2$Cu$_2$(CN)$_3$ in the vicinity of the Fermi energy is 
surprisingly simple. 
Fig.~\ref{FIG:bsdos} shows the band structure and density of states for
$\kappa$-(BEDT-TTF)$_2$Cu$_2$(CN)$_3$. Only two narrow, half filled  bands with 
a width of about 0.4 eV cross the Fermi energy. Those bands are derived almost entirely
from S 3p and C 2p states of the BEDT-TTF molecule and well separated from the remaining 
lower occupied and higher unoccupied bands. About 0.3 to 0.6 eV below the Fermi energy
two further BEDT-TTF derived bands with mainly S 3p and C 2p character are visible.
This is in an overall good agreement with the results of extended H\"uckel 
calculations~\cite{Komatsu96}, 
which find two anti-bonding bands of the highest occupied molecular orbital 
(HOMO) of the BEDT-TTF molecule intersecting the Fermi energy and
place the corresponding bonding bands about 0.4 eV below the
Fermi energy.

\begin{figure}
\includegraphics[angle=-90,width=0.47\textwidth]{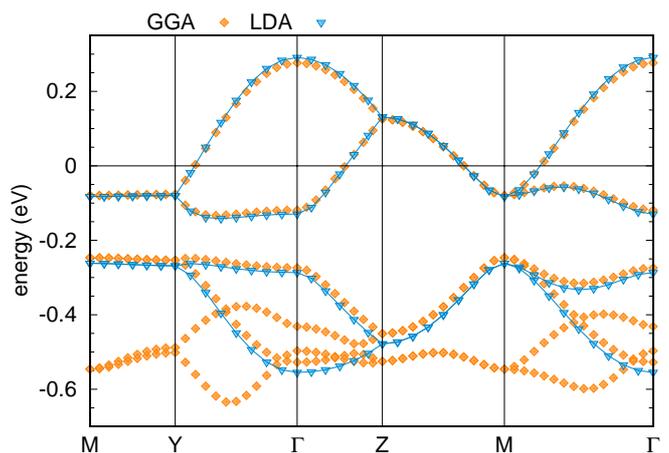}
\caption{Comparison of low energy band structure between different exchange and correlation potentials for $\kappa$-(BEDT-TTF)$_2$X with   X=Cu$_2$(CN)$_3$. The FPLO basis set was used. For GGA, two Cu $3d$ bands are entangled with the bonding bands of the BEDT-TTF dimer molecules; for LDA, they are clearly separate 
(starting below about -0.6 eV) and therefore not shown.
}
\label{FIG:ggavslda}
\end{figure}

\begin{figure}
\includegraphics[angle=-90,width=0.45\textwidth]{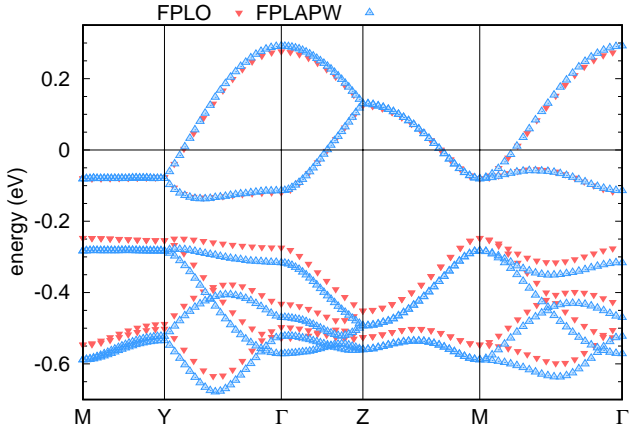}
\caption{Comparison of low energy band structure between FPLAPW and FPLO basis sets for $\kappa$-(BEDT-TTF)$_2$X with   X=Cu$_2$(CN)$_3$. The GGA exchange and correlation potential was used.
}
\label{FIG:fplovswien}
\end{figure}

The Cu 3d states form a narrow band complex about 0.4 to 2.1 eV below the
Fermi energy. The Cu 3d shell is completely
filled, in agreement with the experimental reports~\cite{Komatsu96} that the Cu
atoms are in the Cu$^{1+}$ state with 10 d electrons.
A comparison of the band structure of $\kappa$-(ET)$_2$Cu$_2$(CN)$_3$ near the 
Fermi surface between the GGA and LDA functional is shown in 
Fig.~\ref{FIG:ggavslda}.
The two highest Cu 3d
bands intersect the two bonding BEDT-TTF derived bands in GGA, whereas
the LDA calculations place the Cu 3d bands further below the Fermi energy,
so that the four BEDT-TTF derived bands near the Fermi energy are separated from
the remaining bands within LDA. 
The difference in the position of the Cu 3d level between LDA and GGA
amounts to about 0.2 eV, resulting in a rigid band shift with only
a minute effect on the shape of the BEDT-TTF bands near the Fermi energy and
the model parameters derived from them (see below). 
A similar difference between LDA and GGA derived
band structures is also found in related $\kappa$-(BEDT-TTF)$_2$-X 
compounds~\cite{Kandpal09}.
There are also minor differences in the band structures calculated with the
two different electronic structure codes (Fig.~\ref{FIG:fplovswien}).
While the two BEDT-TTF derived bands crossing the Fermi energy are almost
exactly on top of each other, the lower occupied bands are shifted by
about 50 meV between the FPLO and FPLAPW calculations. A possible reason
for this difference could be the size of the basis set used in the calculations. 
However, we could not analyze the origin of these smaller differences 
in more detail, since the complexity of the crystal structure does not allow 
for extensive numerical checks.

In addition to the
Cu 3d bands there are further BEDT-TTF derived bands in the energy region
between -2.5 and -0.75 eV giving rise to the complex band structure shown 
in Fig.~\ref{FIG:bsdos}. Seperated by a pseudo gap of about 0.5 eV
a complex manifold of BEDT-TTF and anion layer derived bands is
visible starting from 3 eV below the Fermi energy. The higher unoccupied
bands around 2 eV above the Fermi level have predominatly S 3p and C 2p
character.

\begin{figure}
\includegraphics[angle=-90,width=0.47\textwidth]{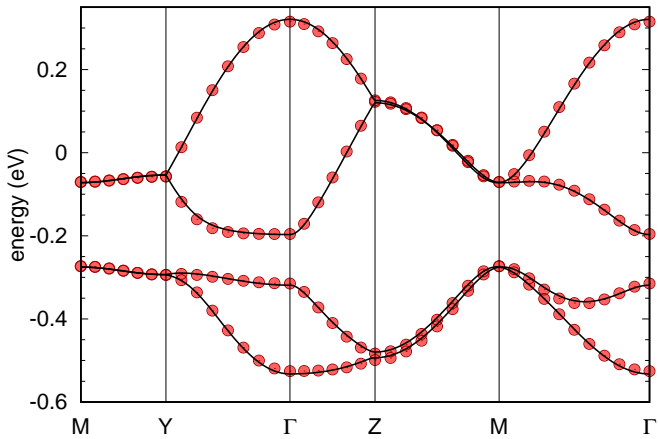}
\caption{Low energy bandstructure of $\kappa$-(BEDT-TTF)$_2$X with   X=Cu(SCN)$_2$ (symbols) and tight binding fit (lines). Note he lifting of the degeneracy at the $Z$ point.
}
\label{SCN_bands}
\end{figure}

As comparison to the bandstructure of  $\kappa$-(ET)$_2$Cu$_2$(CN)$_3$, we present
in Fig.~\ref{SCN_bands} the low energy bandstructure for   $\kappa$-(ET)$_2$Cu(SCN)$_2$.
While the overall features of the two bonding and  antibondig bands are very similar
to those of $\kappa$-(ET)$_2$Cu$_2$(CN)$_3$, $\kappa$-(ET)$_2$Cu(SCN)$_2$ shows a splitting
of the bonding bands at the $Z$ point which is not present in the bandstructure of
 $\kappa$-(ET)$_2$Cu$_2$(CN)$_3$.

\begin{figure}
\includegraphics[width=0.45\textwidth]{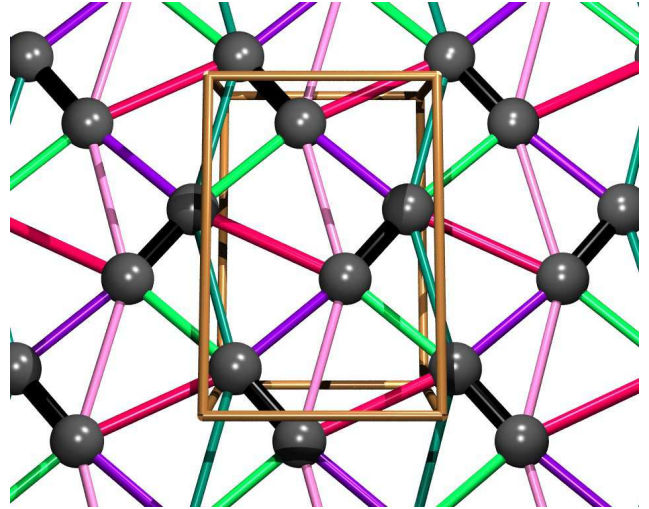}
\caption{Lattice of  BEDT-TTF molecules (grey spheres) for $\kappa$-(BEDT-TTF)$_2$X with   X=Cu(SCN)$_2$. Dimer bonds ($t_1$) are shown in black, $t_4\equiv t'$ in red. $t$ has to be calculated from an average over $t_2$ (light green), $t_3$ (purple), $t_5$ (dark green) and $t_6$ (pink). The view is along the $a$ direction, and the unit cell is marked.
}\label{SCN_structure}
\end{figure}

The lifting of the degeneracy at $Z$ is due to the slightly
distorted arrangement of ET molecules in $\kappa$-(ET)$_2$Cu(SCN)$_2$ as shown
in Fig.~\ref{SCN_structure}.  This feature will also have implications on the determination of the tight-binding parameters
as described below.

\section{Model Hamiltonian}
The  half-filled Hubbard model

\begin{equation}\begin{split}
    H &= \sum_{<ij>,\sigma} t (c_{i\sigma}^\dag
    c_{j\sigma}^{\phantom{\dag}}
    +{\rm H.c.})+\sum_{[ij],\sigma} t' (c_{i\sigma}^\dag c_{j\sigma}^{\phantom{\dag}}+{\rm H.c.})\nonumber\\
    &+U\sum_i \Big(n_{i\uparrow} -\frac{1}{2}\Big)\Big(n_{i\downarrow}
    -\frac{1}{2}\Big)\,.
\label{eq:H}
\end{split}\end{equation}
on the triangular lattice  has been proposed
as the suitable Hamiltonian to describe the behavior of $\kappa$-(BEDT-TTF)$_2$Cu$_2$(CN)$_3$.
$t$ and $t'$, as shown in Fig.~\ref{cn_structure} denote the hopping integrals between nearest neighbors $<i,j>$
and next nearest neighbors $[i,j]$  BEDT-TTF molecule dimers, respectively. Extensions of
this Hamiltonian include also longer ranged hopping integrals, like
$t''$,  as well as nearest neighbor Coulomb integrals. In the 
following we shall concentrate on the calculation of the hopping matrix elements.

In order to obtain the hopping integrals $t$ and $t'$ out of the bandstructure calculations presented above, we have performed various tight-binding
studies. One option, which we denote dimer model,
 consists of defining
a BEDT-TTF dimer of molecules as a lattice site.    The corresponding tight-binding
model is then fitted to the two antibonding bands in Figs.~\ref{FIG:ggavslda}
and \ref{FIG:fplovswien}, and  
the values of $t$, $t'$, $t''$ and longer-ranged hopping parameters are
directly obtained. The
second option, or molecule model,  considers
the molecules in Fig.~\ref{cn_structure} as lattice
sites and the corresponding tight-binding model is fitted to the four
BEDT-TTF derived bands shown in Figs.~\ref{FIG:ggavslda} and 
\ref{FIG:fplovswien}.  The parameters extracted from this
calculation are hopping integrals between molecules. In order
to relate the molecule-molecule hopping parameters with the dimer-dimer
hopping parameters ($t$, $t'$, $t''$ ) we consider the geometrical considerations
given in Ref.~\cite{Komatsu96}. This procedure was done for all bandstructure 
calculations  discussed in the previous section.  In Table~\ref{table_comp}
we present the results for $t$ and $t'$ with the various approaches. 

\begin{table}
\begin{tabular}{llll}
       $t$~(meV)&             $t'$~(meV)&           $t'/t$&      method\\
\hline
\multicolumn{4}{c}{Considering molecules as sites}\\
\hline
        49   & 39   & 0.80 &FPLO(GGA)\\
        50    & 42    & 0.83 &FPLAPW(GGA)\\
        52   & 41   & 0.78 &FPLO(LDA)\\
\hline
\multicolumn{4}{c}{Considering dimers as sites}\\
\hline
        50     & 45   & 0.90 &FPLO(GGA)\\
        52    & 47   & 0.91 &FPLAPW(GGA)\\
        53    & 47   & 0.88 &FPLO(LDA)\\
\hline
\end{tabular}
\caption{Hopping parameters $t$, $t'$ and their ratio $t/t'$ of 
$\kappa$-(ET)$_2$Cu$_2$(CN)$_3$ at ambient pressure, calculated 
with different basis sets (FPLO and FPLAPW) and functionals (LDA and GGA).}
\label{table_comp}
\end{table}

While
the differences between the sets of results are small, they are significant
enough to be relevant for the determination of the $t'/t$ ratio and define
a margin of error of the DFT results. What is important about the results
is that within the margin of error, the DFT calculations provide a clear
prediction for the $t'/t$ $<$ $1$  for   $\kappa$-(ET)$_2$Cu$_2$(CN)$_3$ in disagreement
with extended H\"uckel calculations that estimate  $t'/t$ $>$ $1$.
Recent analysis of angular resolved magnetoresistance oscillations~\cite{Pratt}
 predict
a value of $t'/t$ for $\kappa$-(ET)$_2$Cu$_2$(CN)$_3$ of $\sim 0.93$ in very
good agreement with our results.

In the previous discussion we concentrated on the nearest and next nearest
neighbor hopping integrals $t, t'$. It should be noted though that there are
nonzero longer ranged
hopping terms. In particular $t''$ in the dimer model
has a value of 7~meV ($t''/t=0.14$), which is nonnegligible.

In the tight-binding fit for $\kappa$-(ET)$_2$Cu(SCN)$_2$ the lifting of the degeneracy
at the $Z$ point 
 can only be captured if the $t_2$, $t_3$ and $t_5$, $t_6$ of Fig.~\ref{SCN_structure} are differentiated even though $t_2\approx t_3$ and $t_5\approx t_6$.

Finally we would like to note that while the half-filled Hubbard
model on an anisotropic triangular lattice with $t$ and $t'$ hopping parameters
is the minimal model that we can consider for the description of the system,
a few  extensions may be important like:
(i) inclusion of longer ranged hopping terms, (ii) inclusion of longer ranged
Coulomb interaction terms and (iii) consideration of the system at the level of molecules 
in terms of the quarter-filled Hubbard model~\cite{Li09}.

\section{Conclusions}

In summary we have presented a detailed study of the derivation of Hubbard
model parameters for $\kappa$-(ET)$_2$Cu$_2$(CN)$_3$ in the framework of 
Density Functional Theory. Comparison
of the results obtained with different exchange and correlation functionals,
high precision electronic structure codes as well as different tight
binding models allows us to establish well defined bounds for the 
$t'/t$ ratio used in model calculations. This analysis shows that 
the model parameters for $\kappa$-(ET)$_2$Cu$_2$(CN)$_3$ have to be
qualitatively revised from $t'/t$=1.06~\cite{Komatsu96} to 
$t'/t$=0.83$\pm$0.08. Implications for the properties of
$\kappa$-(ET)$_2$Cu$_2$(CN)$_3$ and possible extensions of the
models are discussed.

\section{Acknowledgements}

We would like to thank M. Lang, J. M\"uller, F. Pratt and J. Merino
 for very useful discussions.
We thank the Deutsche Forschungsgemeinschaft for financial support through
the SFB/TRR 49 and Emmy Noether programs and we would like
to acknowledge support by the Frankfurt Center for Scientific Computing.

\end{document}